\documentclass[a4paper,8pt,twocolumn]{revtex4}

\usepackage[english]{babel}
\usepackage{indentfirst}
\usepackage{graphicx}
\usepackage{amsmath}
\usepackage{amssymb}
\usepackage{amsthm}
\usepackage{mathrsfs}
\usepackage{bbold}
\usepackage{lscape}
\usepackage{bm}
\usepackage{bbm}
\usepackage{float}
\usepackage{longtable}
\usepackage{color}
\usepackage[utf8x]{inputenc}

\def\ket#1{\left|#1\right>}
\def\bra#1{\left<#1\right|}

\def\E#1{\langle #1 \rangle}
\def\eqref#1{Eq.~(\ref{#1})}
\def\figref#1{Fig.~\ref{#1}}
\def\dblone{\hbox{$1\hskip -1.2pt\vrule depth 0pt height 1.6ex width 0.7pt
\vrule depth 0pt height 0.3pt width 0.12em$}}

\newcommand{\kreis}[1]{\unitlength1ex\begin{picture}(2.5,2.5)%
\put(0.75,0.75){\circle{2.5}}\put(0.75,0.75){\makebox(0,0){#1}}\end{picture}}

\begin{document}

\title{Generalized Schrieffer-Wolff Formalism for Dissipative Systems}
\author{E. M. Kessler$^1$}
\affiliation{$^1$ Max-Planck-Institut für Quantenoptik, Hans-Kopfermann-Str. 1
85748 Garching}

\vspace{-3.5cm}

\date{\today}
\begin{abstract}
We present a formalized perturbation theory for Markovian open systems in the language of a generalized Schrieffer-Wolff (SW) transformation. 
A non-unitary rotation decouples the unperturbed steady states from all fast degrees of freedom, in order to obtain an effective Liouvillian, that reproduces the exact low excitation spectrum of the system. The transformation is derived in a constructive way, yielding a perturbative expansion of the effective Liouville operator. The presented formalism realizes an adiabatic elimination of fast degrees of freedom to arbitrary orders in the perturbation.
We exemplarily employ the SW formalism to two generic open systems and discuss general properties of the different orders of the perturbation.


\end{abstract}
 \maketitle

\section{Introduction}
\label{sec:intro}

After more than a century of intensive research, many-body physics is an increasingly thriving field describing most of the phenomena appearing in nature. 
Its goal is to understand the macroscopic properties of large collections of interacting particles (typically of the order of $10^{23}$) from their microscopic laws of motion. 
In typical situations the dynamics of the ensemble is governed by a Hamiltonian $H$ whose complexity and vast dimension impedes a direct solution. However, many complex quantum phenomena  can be understood solely from the low-energy spectrum of $H$, such as quantum phase transitions, topological insulation and superconductivity, just to name a few. 
Therefore a common strategy of many-body-physics is the derivation of a perturbative effective Hamiltonian $H_{\textrm{eff}}$, which approximates the low-energy spectrum of $H$ and reduces the complexity of the problem by integrating out the high-energy degrees of freedom. 
One of the most prominent examples of the success of this approach is the connection between the Kondo model and the low excitation spectrum of the Anderson model, which has been established in 1966 \cite{Schrieffer:1966hu}. It was achieved by a formalized version of (quasi-)degenerate perturbation theory \cite{Klein1974,Shavitt1980,Bravyi:2011fg}, which is nowadays known as \textit{Schrieffer-Wolff} (SW) \textit{transformation} and which paved the way for a deep understanding of these two distinguished models of condensed matter theory \cite{Hewson:1993vc}. 
The many analytical and numerical applications of this perturbative tool in contemporary physics are far too numerous to list here exhaustively (e.g., \cite{MacDonald:1988gr,Paaske:2005gb,Issler2010,Uchoa:2011gt,Lohneysen:2007gm}).

Due to the inevitable coupling of a quantum system to its environment, a paradigm shift could be observed in quantum physics in recent years, as the description of open systems moved into the focus of the field. Many seminal works in the context of, e.g., metrology in the presence of noise \cite{Escher:2011fn,Huelga1997}, dissipative quantum phase transitions \cite{Baumann2010a,Dimer2007}, as well as dissipation assisted quantum state preparation and quantum computation \cite{Kraus:2008jd,Verstraete:2009kc,Shor:jj}, appeared over the past years.
The situation for open systems in many respects parallels the considerations above. For Markovian environments, the system dynamics are described by a non-hermitian Liouville operator $\mathcal L$. 
In many cases one is interested only in the low excitation spectrum of $\mathcal L$, which describes the steady state behavior of the system and comprises in many situations the relevant dynamics, since higher excitations are typically negligibly occupied during the system's evolution.
One prominent example constitutes the emerging field of dissipative phase transitions, which is intimately related to the low excitation spectrum of the Liouvillian \cite{Kessler:2012wi,Morrison2008b,HJCarmichael1980}. 
Also the widespread method of adiabatic elimination of fast evolving degrees of freedom -- routinely employed in the field of quantum optics -- corresponds to the derivation of a perturbative effective Liouvillian describing to low excitation dynamics of a system.
More formalized perturbative tools that accomplish the goal of deriving effective dynamics for open systems to second order have been developed for specific scenarios \cite{Cirac:1992ck,Breuer:2002wp,Reiter2011}.
However, the available tools for open quantum systems are far less advanced than their Hamiltonian analogs. 

In this paper, we present a formalized perturbation theory in the language of a generalized SW transformation, which adapts this formalism for Markovian open systems. 
We consider the most general case of a Liouvillian operator that features an internal hierarchy, i.e., it can be divided into a unperturbed part and a perturbation $\mathcal L = \mathcal L_0 + \epsilon \mathcal V$. 
A non-unitary similarity transformation on $\mathcal L$ dresses the zero eigenstates of $\mathcal L_0$ (i.e., the unperturbed steady states) with higher excitation eigenstates according to the perturbation $\epsilon \mathcal V$ and by construction decouples exactly the corresponding \textit{slow} and \textit{fast space}, respectively.
The projection of the transformed Liouvillian onto this slow space 
(spanned by the dressed steady states of $\mathcal L_0$)
reproduces the exact low excitation spectrum and describes the system evolution in the vicinity of the steady state. 
In analogy to the unitary SW transformation this effective Liouvillian $L_{\textrm{eff}}$ can naturally be expanded in orders of the perturbation parameter $\epsilon$, yielding a systematic perturbative series of the low excitation spectrum and in particular the steady state properties. We stress the point that in contrast to previous perturbative approaches, our formalism works with minimal assumptions on the specific nature of $\mathcal L_0$ and produces perturbative results to arbitrary order.
The procedure formalizes the usual perturbative approach and corresponds to an adiabatic elimination of the fast evolving degrees of freedom. 

The phenomenon that the effective low-energy Hamiltonian derived from integrating out high-energy degrees of freedom, often features a higher complexity than the original one, led in closed systems to the concept of perturbation gadgets \cite{Bravyi:2008bl,Jordan:2008gq,Kempe:2006bj}. 
Along these lines, the idea of \textit{dissipative gadgets}, i.e., the engineering of dissipation for quantum state preparation and protection has recently been proposed \cite{Verstraete:2009kc,Kraus:2008jd,Pastawski:2011ga}. The presented SW formalism provides a natural tool for designing dissipation according to the desired steady state properties. 

The paper is structured as follows. 
In Section~\ref{sec:Formalism} we derive the generalized SW transformation for open systems.
We show, that in the new basis a subspace of slow dynamics decouples exactly from all fast degrees of freedom and we derive an effective Liouvillian within this subspace in a perturbative series [\eqref{eq:Leff}]. 
Subsequently, in Section~\ref{sec:xpls} we employ the formalism in two generic examples, presenting two alternative strategies to evaluate the formal expressions for the effective Liouvillians and discussing general properties of the different orders of the perturbation. 
Finally, in Section~\ref{sec:conclusions} we summarize the results and provide a brief outlook. 

\section{Formalism}
\label{sec:Formalism}

We consider an open system whose evolution is governed by a Markovian master equation. The corresponding Liouville operator can be partitioned in a zeroth order term $\mathcal L_0$ and a perturbation  $\mathcal V$
\begin{align}
\dot\rho=\mathcal L \rho = (\mathcal L_0 + \epsilon \mathcal V)\rho,
\end{align}
where $\epsilon$ denotes the dimensionless perturbation parameter. $\mathcal L_0$ is a linear operator on the vector space of $ \mathbb C^{d\times d}$ matrices ($d$ is the dimension of the system Hilbert space).
We introduce the set of left and right eigenvectors for the non-hermitian operator $\mathcal L_0$.
\begin{align}
\mathcal L_0\ket{r_i} = \lambda_i \ket{r_i},\\
\bra{l_i}\mathcal L_0 = \lambda_i \bra{l_i},
\end{align}
which are chosen to be biorthonormal $\langle l_i | r_j \rangle=\delta_{i,j}$ and generically satisfy the completeness relation $\sum \ket{r_i}\bra{l_i}=\dblone$. The eigenvalues $\lambda_i$ are in general complex.

Since $\mathcal L_0$ is the generator of a universal dynamical map (i.e., a contractive semigroup), its eigenvalues fulfill $\textrm{Re}(\lambda_i)\leq 0$. The generated maps are trace preserving, which guarantees that the kernel of $\mathcal L_0$ is at least one dimensional.
We partition its spectrum in two subsets $\mathcal P=\{\lambda_\alpha | \lambda_\alpha = 0 \}\neq \{\}$ and $\mathcal Q=\{\lambda_i | \lambda_i \neq 0 \}$ (throughout the paper we will refer to eigenvalues from the two sets and the corresponding eigenvectors with greek and arabic indices, respectively). The spectral gap of the unperturbed Liouville operator is denoted as $\Delta = \underset{\lambda_i\in \mathcal Q}{\textrm{min}}(|\lambda_i|)$. 
The in general non-orthogonal projectors 
\footnote{A projector is called orthogonal if its range and null space are orthogonal subspaces.}
\begin{align}
P&=\sum_{\alpha : \lambda_\alpha \in \mathcal P}\ket{r_\alpha}\bra{l_\alpha},\\
Q&=\dblone -P =\sum_{i : \lambda_i \in \mathcal Q}\ket{r_i}\bra{l_i}.
\end{align}
define the subspaces corresponding to the spectral sets $\mathcal P$ and $\mathcal Q$. These subspaces are in the following referred to as the slow (defined by $P$) and fast (defined by $Q$) space, respectively, according to their evolution under the action of $\mathcal L_0$. 

We use this partition of the left and right eigenbases to introduce a block structure for arbitrary superoperators $ A : \mathbb L (\mathcal H) \rightarrow \mathbb L (\mathcal H)$, where $ \mathbb L (\mathcal H)$ denotes the space of linear operators acting on the system's Hilbert space $\mathcal H$ (an example for such a superoperator constitutes the Liouville operator $\mathcal L$ itself),
\begin{align} 
\label{eq:notation}
A=\begin{pmatrix}
A^P  & A^-\\
A^+ & A^Q
\end{pmatrix} =
\begin{pmatrix}
PAP  & PAQ\\
QAP & QAQ
\end{pmatrix}.
\end{align}
Further we introduce block diagonal and block off-diagonal operators
\begin{align}
A^D&=\begin{pmatrix}
A^P  & \mathbb 0\\
\mathbb 0 & A^Q
\end{pmatrix},\\
A^O&=\begin{pmatrix}
\mathbb 0  & A^-\\
A^+ &\mathbb 0
\end{pmatrix}.
\end{align}
By construction the unperturbed Liouville operator is block diagonal in this basis $\mathcal L_0 = \mathcal L_0^D = \mathcal L_0^Q$, while the perturbation in general contains both block diagonal and off-diagonal terms $\mathcal V = \mathcal V^D +  \mathcal V^O$. 
In analogy to the Hamiltonian Schrieffer-Wolff Transformation our goal is to find a similarity transformation 
\begin{align}
\label{eq:simtrafo}
\mathcal L \rightarrow L=U^{-1}\mathcal L U,
\end{align}
such that the two subspaces decouple 
\begin{align}
\label{eq:cond}
L^O=0.
\end{align}

Being similar [\eqref{eq:simtrafo}], the transformed ($L$) and original ($\mathcal L$) Liouvillian share the same spectrum. In the perturbative limit $\Delta > 2 \epsilon ||\mathcal V||$ \cite{Bravyi:2011fg}, the eigenvalues of the superoperator $L_{\textrm{eff}}=PLP$ (referred to as \textit{effective Liouville operator}) reproduce the exact low excitation spectrum of $\mathcal L$. 
The master equation 
\begin{align}
\dot \mu = L_{\textrm{eff}}\mu,
\end{align}
thus describes accurately the steady state properties and low excitation dynamics, i.e., the system evolution in vicinity of the steady state. 
In addition, the Schrieffer-Wolff transformation offers by construction a natural expansion of the effective Liouvillian in the perturbation parameter $\epsilon$. 

In the following we generalize the generic procedure to construct the transformation matrix $U$ for hermitian matrices (for a review see, e.g., \cite{Shavitt1980,Bravyi:2011fg}) to the non-hermitian case. 
It can be shown that \eqref{eq:cond} does not uniquely define the decoupling operator $U$. We will here only consider the so called 'canonical' choice $U=e^S$, where the generator S is imposed to be block off-diagonal $S^D=0$. Other choices of $S^D$ are possible, which then lead to different perturbation theory formalisms, as outlined in \cite{Shavitt1980} for the hermitian case. Depending on the specific problem, alternative gauge choices for $S^D$ may prove to be advantageous. The discussion of the properties of the various formalisms represents an interesting subject for future studies. 

To simplify the formalism we introduce a compact notation where the commutation with an operator $A$ is expressed via the superoperator $\hat A$ defined via:
\begin{align}
\label{eq:not}
\hat A B = [B, A].
\end{align}
Here, $B$ denotes an arbitrary operator. 
This notation allows for an compact representation of the similarity transformation \eqref{eq:simtrafo}
\begin{align}\nonumber
L&=U^{-1}\mathcal L U = e^{-S}\mathcal Le^{S}\\ \label{eq:SSS}
&= \mathcal L + [\mathcal L,S] + \frac{1}{2!} \left[[\mathcal L,S],S\right] + \hdots\\\nonumber
&=\sum_{i=0}^\infty \frac{1}{i!} \hat S^ i\mathcal L = e^{\hat S }\mathcal L.
\end{align}
We partition the latter superoperator into its odd and even powers
\begin{align}
e^{\hat S }=\textrm{cosh}(\hat S) + \textrm{sinh}(\hat S).
\end{align}
The convenience of the canonical choice now becomes evident. While the odd operator $\textrm{sinh}(\hat S)$ changes block diagonal to off diagonal operators and vice versa, the even powers $\textrm{cosh}(\hat S)$ respects that structure. Therefore we can rewrite condition \eqref{eq:cond} 
\begin{align}
\nonumber
L^O = 0& \Leftrightarrow  ( e^{\hat S} \mathcal L )^O=0 \\\label{eq:bdcond}
&\Leftrightarrow \textrm{sinh} (\hat S) \mathcal L^D + \textrm{cosh} (\hat S) \mathcal L^O =0\\\nonumber
&\Leftrightarrow \frac{\textrm{sinh}(\hat S)}{\hat S} \hat S \mathcal L^D + \textrm{cosh} (\hat S) \mathcal L^O =0\\ \nonumber
&\Leftrightarrow    \hat S\mathcal L^D = - \hat S \textrm{coth} (\hat S) \mathcal L^O \\\label{eq:cond2}
&\Leftrightarrow    \hat S\mathcal L_0= - \hat S \epsilon \mathcal V^D - \epsilon \hat S \textrm{coth} (\hat S) \mathcal  V^O,
\end{align}
where in the last step we used $\mathcal L^D = \mathcal L_0 +\epsilon \mathcal V^D$ and $\mathcal L^O=\epsilon \mathcal V^O$.
\eqref{eq:cond2} can be solved formally using resolvent operator techniques. Since by construction the 'slow space' defined by the set $\mathcal P$ contains only unperturbed eigenvalues $\lambda_\alpha=0$, it is $\mathcal L_0 P =P \mathcal L_0=0$ and the resolvent operator takes the simple form
\begin{align}
\mathcal R_0 (A)= Q \mathcal L_0^{-1} A P - P A \mathcal L_0^{-1} Q.
\end{align}
By construction the projection of the zero order Liouvillian into the fast space $Q\mathcal L_0 = \mathcal L_0 Q =Q\mathcal L_0 Q $ has full rank and its inverse is well defined. For simplicity we denote $ \left(Q \mathcal L_0 Q \right)^{-1}\equiv\mathcal L_0^{-1}$.
For block off-diagonal operators $X=X^O$ the resolvent operator fulfills 
\begin{align}
\mathcal R_0 (\hat X \mathcal L_0)=X,
\end{align}
as can be checked straightforwardly.
Applying this superoperator to \eqref{eq:cond2} gives the conditional equation for the generating matrix $S$
\begin{align}
\label{eq:S}
S= -\epsilon \mathcal R_0   \hat S \mathcal V^D -\epsilon \mathcal R_0 \hat S \textrm{coth} (\hat S) \mathcal V^O.
\end{align}

Having derived a formal implicit expression for the transformation matrix $S$, which renders the Liouville operator block diagonal $L^O=0$, we now derive a compact expression for the diagonal blocks $L^D$ in terms of $S$, $\mathcal V$ and $\mathcal L_0$. 
As before, the block off-diagonal structure of $S$ allows us to write $L^D$ as combination of even and odd powers of the superoperator $\hat S$
\begin{align}
L&=L^D=\left(e^{\hat S}\mathcal L\right)^D\\\nonumber
&=\textrm{cosh}(\hat S) \mathcal L^D + \textrm{sinh}(\hat S) \mathcal L^O\\\nonumber
&=\mathcal L^D - \frac{\textrm{cosh}(\hat S)-1}{\textrm{tanh}(\hat S)}\mathcal L^O + \textrm{sinh}(\hat S) \mathcal L^O,
\end{align}
where in the second line we used \eqref{eq:bdcond}.
Using the basic trigonometric relation $\textrm{sinh}(x) - (\textrm{cosh}(x)-1)/\textrm{tanh}(x) = \textrm{tanh}(x/2)$ we find
\begin{align}
L&=\mathcal L^D + \textrm{tanh}(\hat S/2) \mathcal L^O\\\nonumber
&=\mathcal L_0 + \epsilon \left(\mathcal V^D + \textrm{tanh} (\hat S/2)\mathcal V^O\right).
\end{align}
As discussed in \cite{Bravyi:2011fg} all the above hyperbolic transformations are well defined for infinitesimal transformation matrices $S$, which is guaranteed for appropriate perturbation parameters $\epsilon$. We denote the perturbative correction to $\mathcal L_0$ as 
\begin{align}
\label{eq:W}
\mathcal W = \epsilon \left(\mathcal V^D + \textrm{tanh} (\hat S/2)\mathcal V^O\right).
\end{align}
Since by construction $\mathcal L_0P=P\mathcal L_0=0$ the effective Liouville operator  in the slow space is given as
\begin{align}
\label{eq:noname}
L_{\textrm{eff}}=PLP=P\mathcal W P,
\end{align}
and all dynamics in that space are at least of first order in the perturbation. 

Expanding $S$ in orders of the perturbation parameter $\epsilon$ in a Taylor series
\begin{align}
S=\sum_{n=0}^{\infty} \epsilon^n  S_n,
\end{align}
and using \eqref{eq:S} one can deduce a recursive  equation for the $S_n$. With these results we can directly construct the perturbative correction $\mathcal W$ via \eqref{eq:W} order by order. 

The first few Taylor matrices read
\begin{align}
S_0&=0,\\\nonumber
S_1&=- \mathcal R_0 \mathcal V^O= \mathcal V^- \mathcal L_0^{-1} -  \mathcal L_0^{-1} \mathcal V^+, \\\nonumber
S_2& =\mathcal R_0\hat{\mathcal V}^DS_1=-\mathcal R_0\hat{\mathcal V}^D \mathcal R_0 \mathcal V^O,\\\nonumber
\vdots
\end{align}
The corresponding expansion for the perturbative correction matrix $\mathcal W = \sum_{n=0}^{\infty} \epsilon^n  \mathcal W_n  $ can be found via \eqref{eq:W}
\begin{align}
\label{eq:W2}
\mathcal W_0&=0,\\\nonumber
\mathcal W_1&=\mathcal V^D,\\\nonumber
\mathcal W_2&=-\frac{1}{2} \hat{\mathcal V}^O S_1=\frac{1}{2} \hat{\mathcal V}^O R_0 \mathcal V^O,\\\nonumber
\mathcal W_3&=-\frac{1}{2} \hat{\mathcal V}^O S_2=\frac{1}{2} \hat{\mathcal V}^O\mathcal R_0\hat{\mathcal V}^D \mathcal R_0 \mathcal V^O,\\\nonumber
\vdots
\end{align}
In \cite{Bravyi:2011fg} a formal expressions for the $n-th$ order as well as a diagrammatic technique has been derived, which can be directly applied to the case of non-hermitian matrices. 
Straightforward evaluation of Eqs.~(\ref{eq:W2}) and subsequent projection onto $\mathcal P$ yields the first orders of the effective Liouville operator in the slow space [cf. \eqref{eq:noname}]

\begin{align}
\label{eq:Leff}
L_1^{\textrm{eff}}&=P\mathcal V^DP=\mathcal V^P,\\\nonumber
L_2^{\textrm{eff}}&=-P\mathcal V Q \mathcal L_0^{-1} Q\mathcal V P =- \mathcal V^- \mathcal L_0^{-1} \mathcal V^+\\\nonumber
L_3^{\textrm{eff}}&=\mathcal V^-  \mathcal L_0^{-1}  \mathcal V^{Q} \mathcal L_0^{-1} \mathcal V^{+}
- \frac{1}{2} \{ \mathcal V^{P} , \mathcal V^-  \mathcal L_0^{-2}\mathcal V^{+}\}_+,\\\nonumber
\vdots
\end{align}
where we employed the notation introduced in \eqref{eq:notation} and $\{A,B\}_+=AB+BA$ denotes the anti commutator. Note that $\mathcal L^{\textrm{eff}}_2$ reproduces the well know second order result of adiabatic elimination in dissipative systems \cite{Cirac:1992ck}.  

\section{Examples}
\label{sec:xpls}

In this section we will exemplarily employ the formalism developed above in two generic situations and present two alternative strategies to evaluate the expressions for the effective Liouvillians of Eqs.~(\ref{eq:Leff}). 

First, in Section~\ref{sec:gas} we consider the general setting of an ancilla system, which undergoes fast (in general dissipative) dynamics and is weakly coupled to a system. We adiabatically eliminate the ancilla to second order, employing the SW formalism. The coefficients of the effective Liouvillian are expressed in terms of ancilla time correlation functions, which can readily be evaluated using the quantum regression theorem. If one finds a set of operator expectation values with equations of motions that close under the action of $\mathcal L_0$ (in finite dimensional Hilbert spaces this is always the case), the effective master equation can be readily evaluated, even if the ancilla system is high dimensional and its dynamics complicated. We show that the effective Liouville operator to second order is always of Lindblad form \cite{Lindblad1976}, implying a Markovian evolution of the system.

If the zeroth order Liuovillian $\mathcal L_0$ is simple, it is advisable to explicitly calculate the matrices $\mathcal V$ and $\mathcal L_0^{-1}$. Given these matrices, arbitrary orders of the perturbation can readily be evaluated according to \eqref{eq:Leff}. 
In Section~\ref{sec:MS:TO} we consider an example recently studied in the context of superradiance in solid state systems \cite{Kessler:2010fb}, which features simple zeroth order dynamics.
We calculate explicitly the effective Liouvillian up to third order and show that the typically neglected third order has significant impact on the evolution of the system.

\subsection{General Ancilla Setting}
\label{sec:gas}

In the following we consider an example of how to apply the formalism in a generic ancilla setting.
A system is weakly coupled to an (unspecified) ancilla system, which undergoes fast (dissipative and/or coherent) dynamics. 
The Hilbert space of the total system is the product of the ancilla's and system's spaces $\mathcal H = \mathcal H_A \otimes \mathcal H_S$. We assume that the evolution of the ancilla is governed by fast dynamics given by $\mathcal L_0$. $\mathcal L_0$ contains an arbitrary combination of Lindblad and Hamiltonian terms
\begin{align}
\mathcal L_0 \chi= \sum_k \gamma_k (L_k \chi L_k^\dagger - \frac{1}{2} \{L_k^\dagger L_k ,\chi\}_+) - i [H_0,\chi],
\end{align}
where both the $L_k$'s and $H_0$ act only on the ancilla space. Let us for simplicity assume that $\mathcal L_0$ features a unique steady state $\mathcal L_0\sigma_{ss}=0$ such that the projector on the space of zeroth order steady states can be written in the simple form $P\chi=\sigma_{ss} \otimes\textrm{Tr}_A(\chi)\equiv \sigma_{ss} \otimes \mu$ \cite{Zoller:2004wq}. In the last step we introduced the reduced density matrix $\mu\equiv\textrm{Tr}_A(\chi)$.
The weak coupling to the system is realized by the most general Hamiltonian interaction $\epsilon \mathcal V \chi= - i\epsilon [\sum_{\alpha=1}^k A_\alpha \otimes S_\alpha, \chi]$. $A_\alpha$ and $S_\alpha$ are arbitrary hermitian ancilla and system operators, respectively. 
For notational convenience we will suppress the $\otimes$-symbol in the following: $A\otimes S \equiv AS$.
The full master equation thus reads
\begin{align}
\mathcal L \chi= \mathcal L_0 \chi + \epsilon \mathcal V \chi.
\end{align}

Note that the example we consider here corresponds to the often encountered situation of a bipartite system with separation of timescales. The ancilla evolution occurs on a timescale much faster than the system evolution. Thus we can consider the system's evolution under the condition that the ancilla has settled to its steady state. The method presented above represents a formal approach to adiabatically eliminate the fast ancilla dynamics.

\subsubsection{First Order}
The first order in the expansion \eqref{eq:Leff} can readily be evaluated
\begin{align}
L^{\textrm{eff}}_1\chi &= P \mathcal V P\chi =-i \sum_\alpha  P [A_\alpha  S_\alpha,\sigma_{ss}\mu]\\\nonumber
&=-i \sum_\alpha  P \left(A_\alpha \sigma_{ss} [S_\alpha,\mu]   + [A_\alpha, \sigma_{ss} ]S_\alpha \mu \right).
\end{align}
The second term vanishes, since the trace over a commutator is zero: $P [A_\alpha,\sigma_{ss}] S_\alpha \mu=\sigma_{ss}\textrm{Tr}_A([A_\alpha, \sigma_{ss}]) S_\alpha \mu = 0$.
Thus we find the first order of the effective evolution
\begin{align}
L^{\textrm{eff}}_1\chi =\sigma_{ss} L^{\textrm{eff}}_1\mu=- i \sigma_{ss}\sum_\alpha [\E{A_\alpha} S_\alpha,\mu].
\end{align}
Since we are only interested in the evolution of the reduced density matrix $\mu=\textrm{Tr}_A(\chi)$ we can trace out the ancilla degrees of freedom and find the first order correction of the system's evolution
\begin{align}
 L^{\textrm{eff}}_1\mu=- i\sum_\alpha [\E{A_\alpha} S_\alpha,\mu].
\end{align}
Expectedly, to first order, the system experiences merely the effect of the semiclassical values of the ancilla operators.

\subsubsection{Second Order}
The second order of the effective Liouville operator gives rise to more involved dynamics. We calculate the exact expressions and prove its Lindblad form for arbitrary ancilla dynamics $\mathcal L_0$.

For the effective system evolution to second order we have to calculate the expression
\begin{align}
\textrm{Tr}_A (L^{\textrm{eff}}_2 \chi)=-\textrm{Tr}_A(P\mathcal V Q \mathcal L_0^{-1} Q\mathcal V P\chi).
\end{align}
In order to avoid the direct computation of $\mathcal L_0$ which may be impractical for large ancilla systems and for analytical purposes, we express the inverse via the Laplace transform $\mathcal L_0^{-1}=-\int_0^\infty d\tau e^{\mathcal L_0 \tau}$, and we find
\begin{align}
\textrm{Tr}_A (L^{\textrm{eff}}_2 \chi)&= \int_0^\infty d\tau\textrm{Tr}_A(P\mathcal V Q e^{\mathcal L_0 \tau} Q\mathcal V P\chi)\\\nonumber
&=\int_0^\infty d\tau\textrm{Tr}_A\left[P\mathcal V (1-P) e^{\mathcal L_0 \tau} (1-P)\mathcal V P\chi\right]\\\nonumber
&=\int_0^\infty d\tau\textrm{Tr}_A(P\mathcal V e^{\mathcal L_0 \tau} \mathcal V P\chi) \hspace{1cm}\Big\}  ~\kreis{\footnotesize1}\\\nonumber
&\phantom{=}- \int_0^\infty d\tau\textrm{Tr}_A(P\mathcal V  P\mathcal V P\chi), \hspace{1.1cm}\Big\} ~\kreis{\footnotesize2}
\end{align}
where we exploited the property $Pe^{\mathcal L_0 \tau}=e^{\mathcal L_0 \tau}P=P$.

We first evaluate expression $~\kreis{\footnotesize1}$:
\begin{align}
~\kreis{\footnotesize1}&= \int d\tau \textrm{Tr}_A P \mathcal Ve^{\mathcal L_0\tau} \left(-i \left[\sum_j A_j S_j, \sigma_{ss} \mu  \right]\right)\\\nonumber
&=-i\int d\tau \sum_j \textrm{Tr}_A P \mathcal V\left(e^{\mathcal L_0\tau}  A_j \sigma_{ss}\left[ S_j ,  \mu  \right] \right. \\\nonumber
&\hspace{3.5cm}+\left. e^{\mathcal L_0\tau}  \left[A_j , \sigma_{ss}\right] \mu S_j \right)\\\nonumber
&= (-i)^2 \sum_{i,j} \left\{ \int d\tau \textrm{Tr}_A\left( A_i e^{\mathcal L_0 \tau} A_j \sigma_{ss}  \right)\left[ S_i,[S_j,\mu]\right] \right.\\\nonumber
&\hspace{2cm} \left. + \int d\tau \textrm{Tr}_A\left( A_i e^{\mathcal L_0 \tau} [A_j, \sigma_{ss}]  \right)\left[ S_i,\mu S_j\right] \right\}\\\nonumber
&= -  \sum_{i,j} \left\{ \left(\int d\tau\langle A_iA_j(\tau) \rangle_{ss}\right) \left[ S_i,[S_j,\mu]\right] \right.\\\nonumber
&\hspace{1.5cm} \left. + \left(\int d\tau \langle [A_i,A_j(\tau)] \rangle_{ss} \right)  \left[ S_i,\mu S_j\right] \right\}.
\end{align}
In the last step, we defined the time correlation functions in the usual way $\langle A_iA_j(\tau) \rangle_{ss} \equiv \textrm{Tr}_A\left( A_i e^{\mathcal L_0 \tau} A_j \sigma_{ss}  \right) $  and $\langle [A_i,A_j(\tau)] \rangle_{ss} \equiv \textrm{Tr}_A\left( A_i e^{\mathcal L_0 \tau} [A_j ,\sigma_{ss}]  \right) $.
 
 In the same fashion, $~\kreis{\footnotesize2}$ can be readily evaluated to the formal expression
 \begin{align}
~\kreis{\footnotesize2}&=\sum_{i,j} \left(\int_0^\infty d\tau\right)\langle A_i\rangle_{ss}\langle A_j \rangle_{ss} \left[ S_i,[S_j,\mu]\right]. 
\end{align}
In general both formal expressions $~\kreis{\footnotesize1}$ and $~\kreis{\footnotesize2}$ are diverging. However their sum
\begin{align}
\label{eq:secondorder}
&\textrm{Tr}_A (L^{\textrm{eff}}_2 \chi)=~\kreis{\footnotesize1} +  ~\kreis{\footnotesize2}\\\nonumber
&= -  \sum_{i,j} \left\{ \left(\int d\tau\langle \Delta A_i\Delta A_{j\tau} \rangle_{ss}\right) \left[ S_i,[S_j,\mu]\right] \right.\\\nonumber
&\hspace{1.5cm} \left. + \left(\int d\tau \langle [\Delta A_i,\Delta A_{j\tau}] \rangle_{ss} \right)  \left[ S_i,\mu S_j\right] \right\}.
\end{align}
represents a converging and meaningful expression.  We defined $\Delta O_t \equiv O(t) - \E O_{ss}$, for arbitrary ancilla operators $O$.
$~\kreis{\footnotesize2}$ cancels the diverging parts in  $~\kreis{\footnotesize1}$ and renders the integral over correlation functions finite, by subtracting the (infinite) steady state value.

Next, we show that the second order derived above [\eqref{eq:secondorder}] is always of Lindblad form, meaning that it generates a completely positive, trace preserving map. 
According to \eqref{eq:secondorder} the system evolution to second order is entirely determined by the matrix
\begin{align} 
\label{eq:A}
\mathcal A \equiv \left( \int d\tau\langle \Delta A_i\Delta A_{j\tau} \rangle_{ss} \right)_{i,j} = \int d\tau\langle \Delta \vec A\Delta \vec A^*_\tau \rangle_{ss} ,
\end{align}
which can be written as a dyadic product of the vector $\Delta \vec A = (\Delta A_1, \hdots, \Delta A_n)^T$.
In fact, \eqref{eq:secondorder} can be rewritten in the more familiar form
\begin{align}\label{eq:meq22}
&\textrm{Tr}_A (L^{\textrm{eff}}_2 \chi) \\\nonumber
&= \sum_{i,j} \frac{1}{2}(\mathcal A + \mathcal A^\dagger)_{i,j} \left(2 S_j\mu S_i  - \{S_iS_j,\mu\}_+ \right) \\\nonumber
&\hspace{1cm} -i \left[ \frac{1}{2i} \sum_{i,j} (\mathcal A - \mathcal A^\dagger)_{i,j} S_iS_j,\mu \right].
\end{align}

Here it is evident, that the hermitian part of $\mathcal A$ is responsible for the dissipative part of the evolution while the anti-hermitian part defines the coherent evolution. One readily checks that $\frac{1}{2i} \sum_{i,j} (\mathcal A - \mathcal A^\dagger)_{i,j} S_iS_j$ defines a hermitian operator. 
On the other hand, in Appendix~\ref{app:positivity} we show the positivity of the coefficient matrix $\mathcal A + \mathcal A^\dagger \geq 0$, which guarantees Lindblad form of the dissipative term of \eqref{eq:meq22}. Thus, the evolution of the system after adiabatic elimination of the ancilla is  physical and up to second order Markovian. 

Aside from this general result, we now show that the coefficient matrix $\mathcal A$ can readily be calculated without evaluating the respective integrals explicitly, by using the quantum regression theorem \cite{Lax:1963cy}.
Let us assume the equations of motion for the mean deviations of the ancilla operator set $\{A_\alpha\}$ close under $\mathcal L_0$ 
\begin{align}\label{eq:bloch}
\frac{d}{dt}\E{\Delta\vec{A}_t} = \mathcal M \E{\Delta\vec{A}_t}.
\end{align}
In finite dimensional system this can always  be achieved by extending the set $\{A_\alpha |\alpha=1, \hdots, k\}$ to a larger set $\{A_\alpha |\alpha=1, \hdots, n\}$ ($n\geq k$) which forms an operator basis of the ancilla Hilbert space.

Under these conditions the quantum regression theorem allows for a simple evaluation of the relevant time correlation functions
\begin{align}
\frac{d}{dt}&\langle \Delta \vec A_t\Delta \vec A^* \rangle_{ss} = \mathcal M \langle \Delta \vec A_t\Delta \vec A^* \rangle_{ss}\\\nonumber
\Rightarrow&\langle \Delta \vec A_t\Delta \vec A^* \rangle_{ss} = e^{\mathcal M t} \langle \Delta \vec A\Delta \vec A^* \rangle_{ss}.
\end{align}
All eigenvalues of the Bloch matrix $\mathcal M$ have a strictly negative real part (and thus $\mathcal M$ is invertible), since $\mathcal L_0$ generates a contractive semigroup with (by assumption) unique steady state. Therefore the latter equation can be readily integrated yielding a simple expression for the coefficient matrix
\begin{align}
\mathcal A^\dagger = \int_0^{\infty} d\tau\langle \Delta \vec A_\tau\Delta \vec A^* \rangle_{ss} = -\mathcal M^{-1} \langle \Delta \vec A\Delta \vec A^* \rangle_{ss}.
\end{align}
The latter expression can be readily evaluated for a given system and uniquely defines the effective second order dynamics of the system according to Master \eqref{eq:meq22}.
As shown in Appendix~\ref{app:positivity}, independent of the nature and dynamics of the ancilla system, the effective Master \eqref{eq:meq22} is of Lindblad form and gives rise to a Markovian time evolution of the system.

We emphasize that in many situations the size of the minimal set of operators  that close under $\mathcal L_0$ (defining the dimension of $\mathcal M$) will be much smaller than the dimension of the Hilbert space. For illustration, consider the case where the ancilla system is constituted by a driven and damped spin-J and $\{A_\alpha |\alpha=x,y,z\}$ are the usual spin operators. In this case \eqref{eq:bloch} represents optical Bloch equations and the matrix $\mathcal M$ is of dimension 3. Therefore, although the dimension of $\mathcal L_0$ maybe large, the calculation of all coefficients of the effective second order dynamics reduce to a trivial low-dimensional matrix multiplication.

We remark that a similar setting to the one presented here, has recently been examined in the context of dissipative quantum phase transition using the generalized SW technique \cite{Kessler:2012wi}.

\subsection{Mediated Superradiance: Third Order}
\label{sec:MS:TO}

In this Section we examine a specific example of an ancilla setting as discussed above. It recently has been studied in the context of superradiance from nuclear environments of single photon emitters 
\cite{Kessler:2010fb}. A radiatively decaying spin (e.g., a spin pumped electron spin in a quantum dot or nitrogen-vacancy (NV) center \cite{Tamarat:2008ko,Atature:2006ha}) is weakly hyperfine coupled to a large spin environment (e.g., nuclear spins of the host material). After each photon emission the electron spin can escape from the dark state of the dissipation via exchange of an excitation with the nuclear spin environment. Superradiant features in the photon emission originate from a collective enhancement of the hyperfine flip-flop interaction for highly symmetric nuclear states. 

In the following, we derive the effective evolution of the nuclear system after adiabatic elimination of the electron spin up to third order. In contrast to the previous example, where we expressed the effective Liouvillian in terms of integrated time correlation functions (which were evaluated using the quantum regression theorem), we now calculate an explicit matrix representation of the perturbation operator $\mathcal V$, in the biorthonormal eigenbasis of $\mathcal L_0$. Given this representation, all orders can readily be derived by simple matrix multiplication. We will find that the third order in the perturbation significantly improves the accuracy of the perturbative evolution. 

The model we consider is governed by the master equation
\begin{align}
\label{eq:meqex}
\dot \chi = (\mathcal L_0 + \mathcal V) \chi,
\end{align}
where 
\begin{align}
\mathcal L_0 \chi =& \gamma \left( \sigma^- \chi \sigma^+ -\frac{1}{2} \left\{ \sigma^+ \sigma^- ,\chi \right\}_+ \right)\\\nonumber
& - i \omega \left[\sigma^+ \sigma^- ,\chi \right],\\\label{eq:yxc}
\mathcal V \rho =&-i g \left[ \frac{1}{2} \left( \sigma^+ I^- + \sigma^- I^+ \right) +  \sigma^+ \sigma^- I^z, \chi \right],
\end{align}
where $I^\alpha=\sum_{i=1}^N g_i \sigma_i^\alpha$ are collective nuclear spin operators, while $\sigma_i^\alpha$ and $\sigma^\alpha$ are individual nuclear and electronic spin-$1/2$ operators, respectively ($\alpha=\pm,z$). The individual hyperfine coupling constants $g_i$ are normalized to $\sum g_i^2 =1$.
$\gamma$ and $\omega$ denote the photon emission rate of the electron spin and the hyperfine detuning, respectively. The nuclear and electronic system are weakly hyperfine coupled with $g \sqrt N\ll \gamma, \omega$. 
This model describes the superradiant evolution of an NV center coupled to a nuclear spin environment, as discussed in \cite{Kessler:2010fb}. 

After the assignment $\chi =  \sum_{i,j} \chi_{i,j} \ket i \bra j \rightarrow \vec \chi = \sum_{i,j} \chi_{i,j} \ket i \otimes \ket j$ (where $\{\ket i \bra j\} $ is an arbitrary basis of the matrix vector space), the
superoperators of \eqref{eq:meqex} can be written in matrix representation as
\begin{align}
\mathcal L_0 =& \gamma\left[ \sigma^- \otimes (\sigma^+)^T - \frac{1}{2} \left(\sigma^+ \sigma^- \otimes \dblone + \dblone \otimes (\sigma^+ \sigma^-)^T \right)\right]\\\nonumber
& - i \omega \left( \sigma^+ \sigma^- \otimes \dblone - \dblone \otimes (\sigma^+ \sigma^-)^T \right),
\end{align}
and
\begin{align}
\mathcal V = - ig&\left[ \left(\frac{1}{2} \left( \sigma^+ I^- + \sigma^- I^+ \right) +  \sigma^+ \sigma^- I^z  \right) \otimes \dblone \right. \\\nonumber
&\left. - \dblone \otimes  \left(\frac{1}{2} \left( \sigma^+ I^- + \sigma^- I^+ \right) +  \sigma^+ \sigma^- I^z  \right)^T \right],
\end{align}
where the superscript $T$ denotes the matrix transpose. Since $\mathcal L_0$ acts only on the electronic space, it can be straightforwardly diagonalized in the basis of left and right eigenvectors
\begin{align}
\mathcal L_0=\sum_{i=1}^4 \lambda_i \ket{r_i} \bra{l_i}.
\end{align}
The eigenvalues and the biorthonormal ($\langle l_i | r_j \rangle =\delta_{i,j}$) left and right eigenvectors are given in Table~\ref{tab:1}.

\begin{centering}
\begin{table}
\renewcommand{\arraystretch}{1.5}
    \begin{tabular}{|c||c|c|c|l}\hline
 & $\lambda_i$ & $\ket{r_i}$ & $\ket{l_i}$ \\
 \hline \hline
  \#1&$ \lambda_1=0$ &$\ket{r_1} = \ket{\downarrow \downarrow}$ &$ \ket{l_1} = \ket{\downarrow \downarrow} + \ket{\uparrow \uparrow}   $ \\
 \hline
\#2&$ \lambda_2=-\gamma/2 +i\omega$ &$\ket{r_2} = \ket{\downarrow \uparrow}$ &$ \ket{l_2} = \ket{\downarrow \uparrow}   $ \\
 \hline
  \#3 &$ \lambda_3=-\gamma/2 -i\omega$ &$\ket{r_3} = \ket{\uparrow \downarrow}$ &$ \ket{l_3} =  \ket{\uparrow \downarrow}  $ \\
 \hline
\#4 &$ \lambda_4=-\gamma$ &$\ket{r_1} =\ket{\uparrow \uparrow}- \ket{\downarrow \downarrow}$ &$ \ket{l_4} = \ket{\uparrow \uparrow}   $ \\
\hline
    \end{tabular}
    \caption{Eigenvalues and left and right eigenvectors of $\mathcal L_0$. We used the simplified notation $\ket{i j} \equiv \ket i \otimes \ket j $ ($i,j = \uparrow, \downarrow$), which are the vector representation of the basis matrices $\ket i \bra j$ of the electronic space (cf. text).  
    } 
    \label{tab:1}
\end{table}
\end{centering}

The representation of the perturbation in this basis 
\begin{align}\label{eq:Vop}
\mathcal V&= \sum_{i,j=1}^4 \mathcal V_{i,j} \ket{r_i} \bra{l_j},\\
\mathcal V_{i,j}& = \bra{l_i} \mathcal V \ket{r_j},
\end{align}
can readily be derived and is given as 
\begin{widetext}
\begin{align}
\label{eq:V}
\mathcal V&=
\left(
	\begin{array}{c|ccc}
	\renewcommand{\arraystretch}{2}
		 \mathcal V^P & &\mathcal V^- & \\	 
		\hline&&&  \\
		 \mathcal V^+ &&\mathcal V^Q& \\
 		&&&
	\end{array}\right)\\\nonumber
&=	
\left(
	\begin{array}{c|ccc}
	\renewcommand{\arraystretch}{2}
		 0 & -i g/2 \left(I^- \otimes \dblone - \dblone \otimes (I^-)^T\right) & -i g/2 \left(I^+ \otimes \dblone - \dblone \otimes (I^+)^T \right)  &-i g \left(I^z \otimes \dblone - \dblone \otimes (I^z)^T\right) \\	 
		\hline ig/2 ~ \dblone \otimes (I^+)^T&  ig ~ \dblone \otimes (I^z)^T & 0 & -i g/2 \left(I^+ \otimes \dblone + \dblone \otimes (I^+)^T \right) \\
		  -ig/2 ~ I^- \otimes\dblone & 0 & -ig ~ I^z \otimes\dblone & i g/2 \left(I^- \otimes \dblone + \dblone \otimes (I^-)^T\right) \\
 		 0 &  -ig/2 ~ I^- \otimes\dblone  & ig/2 ~ \dblone \otimes (I^+)^T & -ig  \left(I^z \otimes\dblone - \dblone \otimes (I^z)^T\right) 
	\end{array}\right).
\end{align}
\end{widetext}
Note that for simplicity we denote operators [e.g., \eqref{eq:Vop}] and their representation in the $\mathcal L_0$ eigenbasis [e.g., \eqref{eq:V}] with the same symbols. 
The inverse of $\mathcal L_0$ in the fast space is simply given as 
\begin{align}\label{eq:Linv}
 \mathcal L_0^{-1} \equiv \left(Q\mathcal L_0 Q \right)^{-1}=
\left(
	\begin{array}{ccc}
	\renewcommand{\arraystretch}{2}
		 1/\lambda_2 &0 &0  \\	 
		 0&1/\lambda_3& 0 \\
		 0&0&1/\lambda_4  \\
	\end{array}\right).
\end{align}
All orders of the perturbation can now readily be derived from products of the above matrices \eqref{eq:V} and \eqref{eq:Linv}, according to \eqref{eq:Leff}. 
In the following we calculate explicitly the first three orders of the effective nuclear Liouville operator $L^{\textrm{eff}}$. The first order $L^{\textrm{eff}}_1$ vanishes, since the perturbation vanishes in the slow space, $\mathcal V^P=0$. 
The second order yields 
\begin{align}
\label{eq:2nd}
L^{\textrm{eff}}_2 &= -\mathcal V^- \mathcal L_0^{-1} \mathcal V^+\\\nonumber
=&\gamma_{\textrm{eff}} \left[I^-\otimes (I^+)^T - \frac{1}{2} \left( \dblone \otimes (I^+I^-)^T + I^+I^-\otimes \dblone \right)  \right]\\\nonumber
& -i \omega_{\textrm{eff}} \left[ I^+I^-\otimes \dblone - \dblone \otimes (I^+I^-)^T \right],
\end{align}
where $\gamma_{\textrm{eff}} = -2 \textrm{Re}(g^2/\lambda_2) = g^2 \gamma/[(\gamma/2)^2 + \omega^2]$ and $\omega_{\textrm{eff}} =  \textrm{Im}(g^2/\lambda_2) = -g^2 \omega /[(\gamma/2)^2 + \omega^2]$.

In the standard representation this corresponds to the second order master equation
\begin{align}
\label{eq:2ndst}
\dot \mu = L^{\textrm{eff}}_2\mu= &\gamma_{\textrm{eff}} \left[ I^- \mu I^+ -\frac{1}{2} \left\{ I^+ I^- ,\mu \right\}_+ \right] \\\nonumber
&- i \omega_{\textrm{eff}} \left[I^+ I^- ,\mu \right],
\end{align}
which describes the collective decay of the nuclear spins at rate $\gamma_{\textrm{eff}}$, responsible for a superradiant evolution as discussed in \cite{Kessler:2010fb}. It is of Lindblad form and agrees with the result derived using standard adiabatic elimination techniques. 

After having derived the matrix representation of $\mathcal V$ [\eqref{eq:V}] the SW formalism allows us to readily evaluate higher order corrections according to \eqref{eq:Leff}. This contrasts the situation for the standard techniques of adiabatic elimination, where the derivation of higher order terms is extremely tedious. 
The third order of \eqref{eq:Leff} yields the more involved expression (note that $\mathcal V^P=0$)
\begin{align}
\label{eq:3rd}
L^{\textrm{eff}}_3=& \mathcal V^-  \mathcal L_0^{-1}  \mathcal V^{Q} \mathcal L_0^{-1} \mathcal V^{+}\\\nonumber
=&i \frac{g^3}{4\lambda_2^2} \left[I^- \otimes (I^+I^z)^T - \dblone \otimes (I^+I^zI^-)^T     \right] \\\nonumber
& - i \frac{g^3}{4\lambda_2 \lambda_4} \left[I^zI^- \otimes (I^+)^T - I^- \otimes (I^+I^z)^T     \right]\\\nonumber
& + i \frac{g^3}{4\lambda_3^2} \left[I^+I^zI^- \otimes \dblone - I^zI^- \otimes (I^+)^T     \right]\\\nonumber
& - i \frac{g^3}{4\lambda_3 \lambda_4} \left[I^zI^- \otimes (I^+)^T - I^- \otimes (I^+I^z)^T     \right],
\end{align}
All terms involve contributions of the term $\propto \sigma^+\sigma^- I_z$ in \eqref{eq:yxc}. It enters via $\mathcal V^Q$ in the intermediate process of \eqref{eq:3rd}. In contrast, the second order was entirely independent of this term.

\begin{figure}[ht]
\centering
\includegraphics[width=0.45\textwidth]{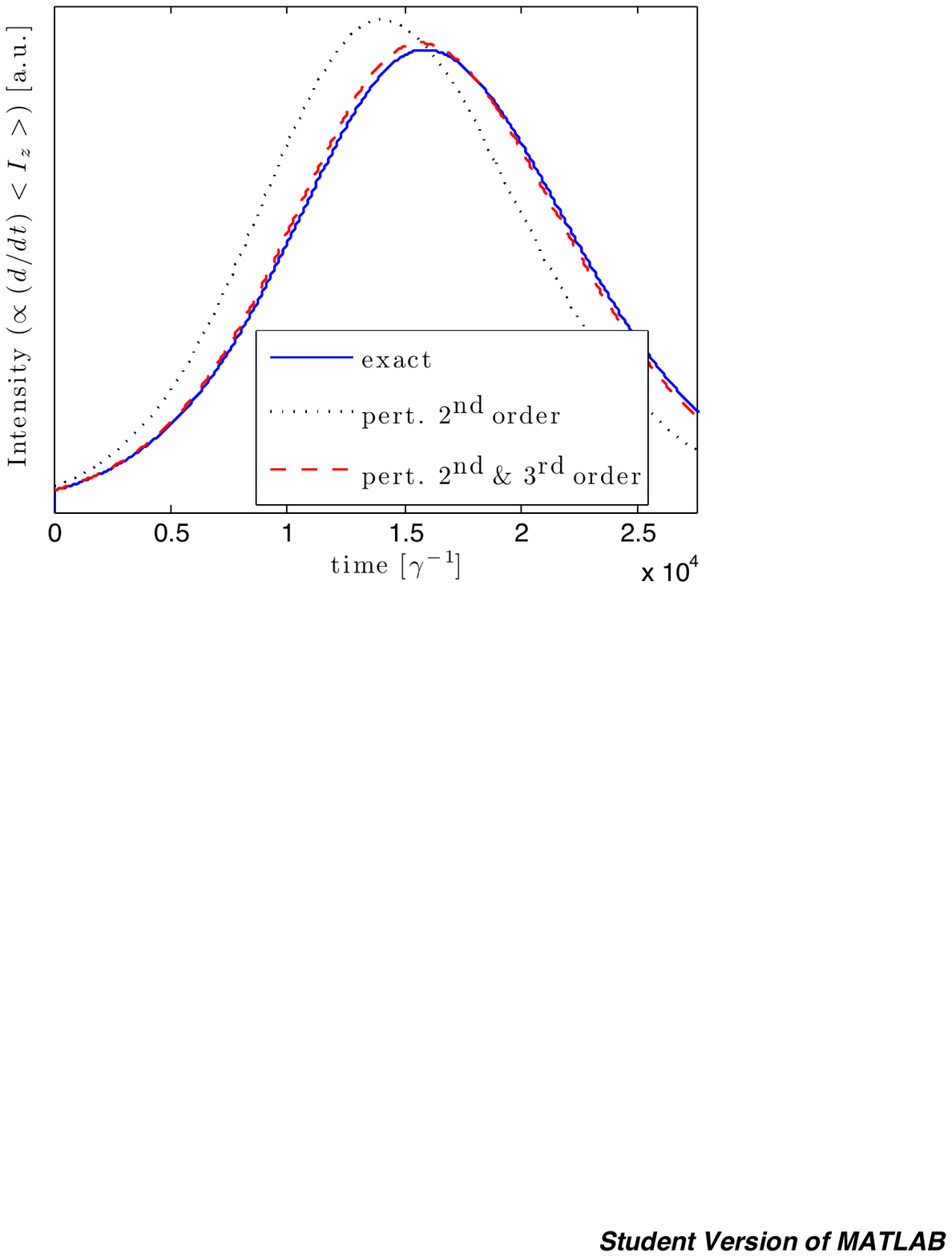}
  \caption{Comparison of the exact evolution according to \eqref{eq:meqex} (dashed) with the approximate solution up to second (dotted) and third (solid) order [\eqref{eq:2nd} \& \eqref{eq:3rd}]. The figure shows the radiation intensity emitted from the electron spin ($\propto d/dt \E{I_z}$) for a system of $N=100$ nuclear spins, initially fully polarized in $z$ direction. The emission intensity shows clearly the characteristic superradiant burst \cite{Kessler:2010fb}. Parameters are $\omega=\gamma/5$ and $\sqrt N g = 0.2 \gamma$. The second order effective evolution ($L^{\textrm{eff}}_2$) shows significant deviation from the exact dynamics. Addition of the third order $L^{\textrm{eff}}_3$, improves the approximation substantially.}  \label{fig:1}
\end{figure}

In \figref{fig:1} we compare simulations of the exact evolution according to \eqref{eq:meqex} (solid) with the perturbative solution up to second (dotted) and third (dashed) order, respectively, for a system of $N=100$ homogeneously coupled nuclear spins. The spins are initially fully polarized in $z$ direction. 
We find that the third order effective Liouvillian has significant impact on the accuracy of the approximation.

It can be shown that the Liouvillian \eqref{eq:3rd} is hermiticity and trace preserving. Furthermore, the density matrix of the system remains positive throughout the evolution, although the third order Liouvillian it is not obviously of Lindblad form. In the following we show that in the limit $\omega=0$ one can recover Lindblad form by adding terms of higher order in the perturbation. In this case, all non-Lindblad effects are of an order that is deliberately neglected.

For $\omega=0$, \eqref{eq:3rd} reduces to the simple expression (in standard representation)
\begin{align}
\label{eq:}
L^{\textrm{eff}}_3\mu=  i \frac{g}{2\gamma}  \gamma_{\textrm{eff}} [ I^-\mu I^+, I_z ] + i \frac{g}{4\gamma}\gamma_{\textrm{eff}} [ I^+ I^z I^-, \mu ].
\end{align}

The first term of $L^{\textrm{eff}}_2$ can be combined with the first term of $L^{\textrm{eff}}_3$
\begin{align}
\label{eq:}
&\gamma_{\textrm{eff}} \left( I^- \mu I^++ i \frac{g}{2\gamma}    [ I^-\mu I^+, I_z ] \right) \\\nonumber
=& \gamma_{\textrm{eff}} \left( e^{-i \frac{g}{2\gamma}I^z }I^- \mu I^+e^{i \frac{g}{2\gamma}I^z } + O\left[(g/\gamma)^2\right]\right),
\end{align}
where we used relation \eqref{eq:SSS}. The term $O\left[(g/\gamma)^2\right]$ is of fourth order in the perturbation and can consistently be neglected.
The resulting master equation up to third order then has Lindblad form
\begin{align}
\label{eq:}
&\left(L^{\textrm{eff}}_2 + L^{\textrm{eff}}_3\right)\mu \\\nonumber
=& \gamma_{\textrm{eff}} \left[e^{-i \frac{g}{2\gamma}I^z }  I^- \mu I^+e^{i \frac{g}{2\gamma}I^z }  -\frac{1}{2} \left\{ I^+ I^- ,\mu \right\}_+ \right]\\\nonumber
    &+ i \frac{g}{4\gamma}\gamma_{\textrm{eff}} [ I^+ I^z I^-, \mu ].    
\end{align}

The excellent agreement of perturbative and exact solution displayed in \figref{fig:1} supports the expectation that similar arguments hold in the general case $\omega \neq 0$ [\eqref{eq:3rd}] and effects due to the non-Lindblad form of the effective Liouvillian are of higher order in the perturbation. This assumption as well as the possibility to use the gauge invariance under the choice of $S^D$, in order to ensure Lindblad form of every order of the perturbation, is subject to future work. 

\section{Conclusions}
\label{sec:conclusions}

We presented a generalized SW formalism, which adapts the successful perturbative tool of Hamiltonian quantum mechanics to the case of open quantum systems, whose evolution is governed by a Liouville operator. 
In analogy to the coherent case, we derive a transformation, that decouples subspaces of slow dynamics from fast evolving degrees of freedom in a perturbative series.
In comparison with alternative schemes for adiabatic elimination \cite{Reiter2011,Breuer:2002wp,Cirac:1992ck}, the advantages of the presented method are twofold. First, minimal assumptions on the specific type of the zero order dynamics $\mathcal L_0$ have to be made. The subspace to be eliminated can be high dimensional and undergoing involved dynamics. Second, our approach in principle is an exact decoupling scheme, and can be applied to all orders in the perturbation. This property is of particular relevance for instance in numerical studies of low excitation spectra of the Liouville operator, e.g., in the context of dissipative phase transitions \cite{Kessler:2012wi,Morrison2008b,HJCarmichael1980}, and for error estimation in the context of quantum information processing \cite{Pastawski:2011ga}. The SW formalism provides a natural framework for the engineering of dissipative gadgets, e.g., in the context of state preparation and protection \cite{Verstraete:2009kc,Kraus:2008jd,Pastawski:2011ga}.

We employed the SW formalism exemplarily to two model systems and presented different schemes to evaluate the expressions for the effective Liouvillians.
In a generic ancilla setting, we proved that the effective evolution of the weakly coupled system is up to second order Markovian, irrespective of the specific realization of the ancilla. In a second example we demonstrated that - in contrast to the standard schemes of adiabatic elimination - higher order corrections can readily be derived within the SW framework. In this model, the third order correction plays a significant role in the perturbative dynamics.  

Further, we point out that the freedom of a gauge choice in the derivation of the transformation matrix could lead to a set of alternative perturbative approaches, which, in analogy to the standard SW transformation and depending on the specific problem, could prove to be advantageous under certain conditions. Potentially, this gauge freedom could also be used to ensure Lindblad form of the higher order effective Liouvillians. 
Lastly, we mention the numerous theoretical results \cite{Bravyi:2011fg} (e.g., linked cluster theorem, additivity of effective Hamiltonian), which have been derived in the context of coherent SW transformations as well as different variations of the SW method (e.g., continuous SW \cite{Kehrein:tv}), which may have open system analogs. These questions will be subject to future studies. 

\acknowledgements
We acknowledge support by the DFG within SFB 631 and the Cluster of Excellence NIM. Further, we thank Geza Giedke and Ignacio Cirac for many fruitful discussions.

\appendix
\section{Lindblad Form of \eqref{eq:meq22}}
\label{app:positivity}

In the following, we prove that the dissipative part of the second order effective Liouvillian \eqref{eq:meq22} is of Lindblad form. For this we have to show positivity of the respective coefficient matrix
\begin{align}\label{eq:A1}
\mathcal A + \mathcal A^\dagger &\geq 0 \\\nonumber
\Leftrightarrow \vec v^* (\mathcal A + \mathcal A^\dagger) \vec v &\geq 0, \hspace{1cm} \forall \vec v \in \mathbb C^n
\end{align}

Expressing $\mathcal A$ as the integrated dyadic product [\eqref{eq:A}] we write
\begin{align}
\label{eq:pos}
\vec v^* &(\mathcal A + \mathcal A^\dagger) \vec v \\\nonumber
&=\int_0^\infty d\tau  \left( \langle \vec v^* \Delta \vec A\Delta \vec A^*_\tau \vec v \rangle_{ss} +  \langle\vec v^* \Delta \vec A_\tau\Delta \vec A^*\vec v \rangle_{ss} \right)\\\nonumber
&=\int_0^\infty d\tau  \left( \langle\sigma \sigma_\tau^\dagger \rangle_{ss} +  \langle\sigma_\tau \sigma^\dagger \rangle_{ss} \right),
\end{align} 
where we introduced the ancilla operator $\sigma \equiv \vec v^*\Delta \vec A$. 

Since the expectation values are evaluated in the ancilla's steady state, the two-time correlation functions are invariant under a total time translation $t$:
\begin{align}\label{eq:tsym}
 \langle\sigma \sigma_\tau^\dagger \rangle_{ss} =  \langle\sigma_t \sigma_{t+\tau}^\dagger \rangle_{ss}\\
 \langle\sigma_\tau \sigma^\dagger \rangle_{ss} =  \langle\sigma_{t+\tau} \sigma_{t}^\dagger \rangle_{ss}.
\end{align}

We exploit that property in symmetrizing \eqref{eq:pos} in the time arguments. First we 'average' \eqref{eq:pos} over a total time translation
\begin{align}
\vec v^* &(\mathcal A + \mathcal A^\dagger) \vec v \\\nonumber
 &=\frac{1}{t_0}\int_0^{t_0}dt \int_0^\infty d\tau  \left( \langle\sigma_t \sigma_{t+\tau}^\dagger \rangle_{ss} +  \langle\sigma_{t+\tau} \sigma_t^\dagger \rangle_{ss} \right)\\\nonumber
  &=\frac{1}{t_0}\int_0^{t_0}dt \int_{t}^\infty dt'  \left( \langle\sigma_t \sigma_{t'}^\dagger \rangle_{ss} +  \langle\sigma_{t'} \sigma_t^\dagger \rangle_{ss} \right),
\end{align}
where the new variable $t'=t+\tau$ has been introduced.
Basic integral transformations lead to the expression
\begin{align}
\vec v^* &(\mathcal A + \mathcal A^\dagger) \vec v \\\nonumber
  =&\frac{1}{t_0}\int_0^{t_0}dt \int_0^{t_0} dt'   \langle\sigma_t \sigma_{t'}^\dagger \rangle_{ss}\hspace{1.1cm}\Big\} \hspace{0.3cm}\kreis{\footnotesize a} \\\nonumber
  &+\frac{1}{t_0}\int_0^{t_0}dt \int_{t_0}^{\infty} dt'   \langle\sigma_t \sigma_{t'}^\dagger \rangle_{ss} \hspace{1.1cm}\Big\} \hspace{0.3cm}\kreis{\footnotesize b}
\\\nonumber
  &+\frac{1}{t_0}\int_0^{t_0}dt' \int_{t_0}^{\infty} dt   \langle\sigma_t \sigma_{t'}^\dagger \rangle_{ss} .\hspace{1.1cm}\Big\} \hspace{0.3cm}\kreis{\footnotesize c}
\end{align}

The first term of the latter equation (~\kreis{\footnotesize a}) is positive since 
\begin{equation}
\kreis{\footnotesize 1} = \E{R R^\dagger}_{ss} \geq 0,
\end{equation}
with $R=\left(1/\sqrt t_0\right) \int_{0}^{t_0} dt  ~ \sigma_t $. 

We show that the remaining terms vanish in the limit $t_0\rightarrow \infty$, proving the Lindblad form of \eqref{eq:secondorder}. We estimate
\begin{align}
\left |~\kreis{\footnotesize b}\right |& =\left | \frac{1}{t_0}\int_0^{t_0}dt \int_{t_0-t}^{\infty} d\tau \E{\sigma \sigma_\tau^\dagger}_{ss}\right | \\\nonumber
 &\leq  \frac{1}{t_0} \int_0^{t_0}dt \int_{t_0-t}^{\infty} d\tau\left | \E{\sigma \sigma_\tau^\dagger}_{ss}\right | ,
\end{align}
where we reintroduced the time difference integration variable $\tau = t' - t$ and used the time translation symmetry \eqref{eq:tsym}. Next the integration over $dt$ is divided into two parts defined by the parameter $x$: $\int_0^{t_0}dt = \int_0^{t_0-x}dt + \int_{t_0-x}^{t_0}dt$. The first term can be upper bounded as
\begin{align}\label{eq:a1}
\frac{1}{t_0} \int_0^{t_0-x}&dt \int_{t_0-t}^{\infty} d\tau\left | \E{\sigma \sigma_\tau^\dagger}_{ss}\right | \\\nonumber
 &\leq  \frac{1}{t_0} \int_0^{t_0-x}dt \int_{x}^{\infty} d\tau\left | \E{\sigma \sigma_\tau^\dagger}_{ss}\right | \\\nonumber
 &\leq  \int_{x}^{\infty} d\tau\left | \E{\sigma \sigma_\tau^\dagger}_{ss}\right | .
\end{align}

The second term can be estimated as
\begin{align}\label{eq:a2}
\frac{1}{t_0} \int_{t_0-x}^{t_0}&dt \int_{t_0-t}^{\infty} d\tau\left | \E{\sigma \sigma_\tau^\dagger}_{ss}\right | \\\nonumber
\leq &\frac{1}{t_0} \int_{t_0-x}^{t_0}dt \int_{0}^{\infty} d\tau\left | \E{\sigma \sigma_\tau^\dagger}_{ss}\right | \\\nonumber
= &\frac{x}{t_0} \int_{0}^{\infty} d\tau\left | \E{\sigma \sigma_\tau^\dagger}_{ss}\right |.
\end{align}
Using the quantum regression theorem one shows that the time correlation function $\E{\sigma \sigma_\tau^\dagger}_{ss}$ decays exponentially.
Choosing the parameter $x=\sqrt t_0$ both the right hand side of \eqref{eq:a1} and \eqref{eq:a2} vanish in the limit $t_0\rightarrow \infty$. In an analogous estimation one shows the vanishing of the remaining term ~\kreis{\footnotesize c} which proves the positivity \eqref{eq:A1} and thus the Lindblad form of \eqref{eq:meq22}.


\end{document}